\documentclass[journal=jacsat,manuscript=article]{achemso}

\usepackage[version=3]{mhchem} 

\author{Pavel Tonkaev}
\affiliation{Hybrid Nanophotonics and Optoelectronics Laboratory, Physics and Engineering Department, ITMO University, St.Petersburg, 197101, Russia}
\author{George Zograf}
\affiliation{Hybrid Nanophotonics and Optoelectronics Laboratory, Physics and Engineering Department, ITMO University, St.Petersburg, 197101, Russia}
\author{Sergey Makarov}
\affiliation{Hybrid Nanophotonics and Optoelectronics Laboratory, Physics and Engineering Department, ITMO University, St.Petersburg, 197101, Russia}
\email{s.makarov@metalab.ifmo.ru}

\title  {Optical cooling of lead halide perovskite nanoparticles enhanced by Mie resonances}

\abbreviations{IR,NMR,UV}

\begin{document}

\section{\label{sec:level1} Introduction}

Halide organometal perovskite materials attracted a major attention owing to their outstanding photovoltaic properties and performance~\cite{snaith2018present}. Indeed, halide perovskite-based single-junction solar cells demonstrated increase of efficiency from 3.8\%~\cite{kojima2009organometal} to over 21\%~\cite{bi2016polymer, green2018solar} in under 10 years period. Efficiency and performance of perovskite based light-emitting devices are also rapidly growing~\cite{cao2018perovskite}, which make them prospective for applications in displays. However, halide perovskites suffer from relatively low thermal conductivity~\cite{pisoni2014ultra}, which results in devices overheating and might strongly affect their applicability. Thus, finding the way for efficient cooling of perovskite active layers in devices would solve number of technological problems.

One of the possible solutions is optical cooling, when absorption of photon with lower energy than that of emitted (so-called up-conversion) results in absorption of additional phonon for energy conservation. Low trap-state density~\cite{shi2015low} and high photoluminescence quantum yield~\cite{richter2016enhancing}, and pronounced excitonic states at room temperature in bulk halide perovskites resulted in high efficiency of up-conversion~\cite{ha2016laser}, and, thus, in the decrease of local temperature by 20~K upon laser irradiation. Further optimization of the optical cooling requires both enhanced quantum yield of emission and improved absorption of incident light in the material.
Therefore, nanophotonic structures with optical resonances can be beneficial in terms of optical cooling. The perovskites also have a fairly high refractive index (2.0 - 2.5 around exciton), which makes them promising for using in nanophotonics applications due to the ability of a single dielectric resonant nanoparticle (NP) to achieve strong resonant response and light localization~\cite{kuznetsov2016optically}, whereas the excitation of resonances in the nanocavities results in sufficient improvement of emission rate~\cite{purcell1946purcell}. 
The combination of nanophotonics designs~\cite{makarov2019halide} with the variety of functional properties of the halide perovskites~\cite{sutherland2016perovskite} already demonstrated great performance in photoluminescence efficiency boost from metasurfaces~\cite{gholipour2017organometallic, tiguntseva2017resonant,tiguntseva2018light}, tunability of luminescence~\cite{gao2018lead} and optical resonances~\cite{tiguntseva2018tunable}, and outstanding properties of microlasers~\cite{shi2015low, du2018recent, zhizhchenko2019single}.

\begin{figure}[ht!]
    \centering
   
   \includegraphics[width=1\linewidth]{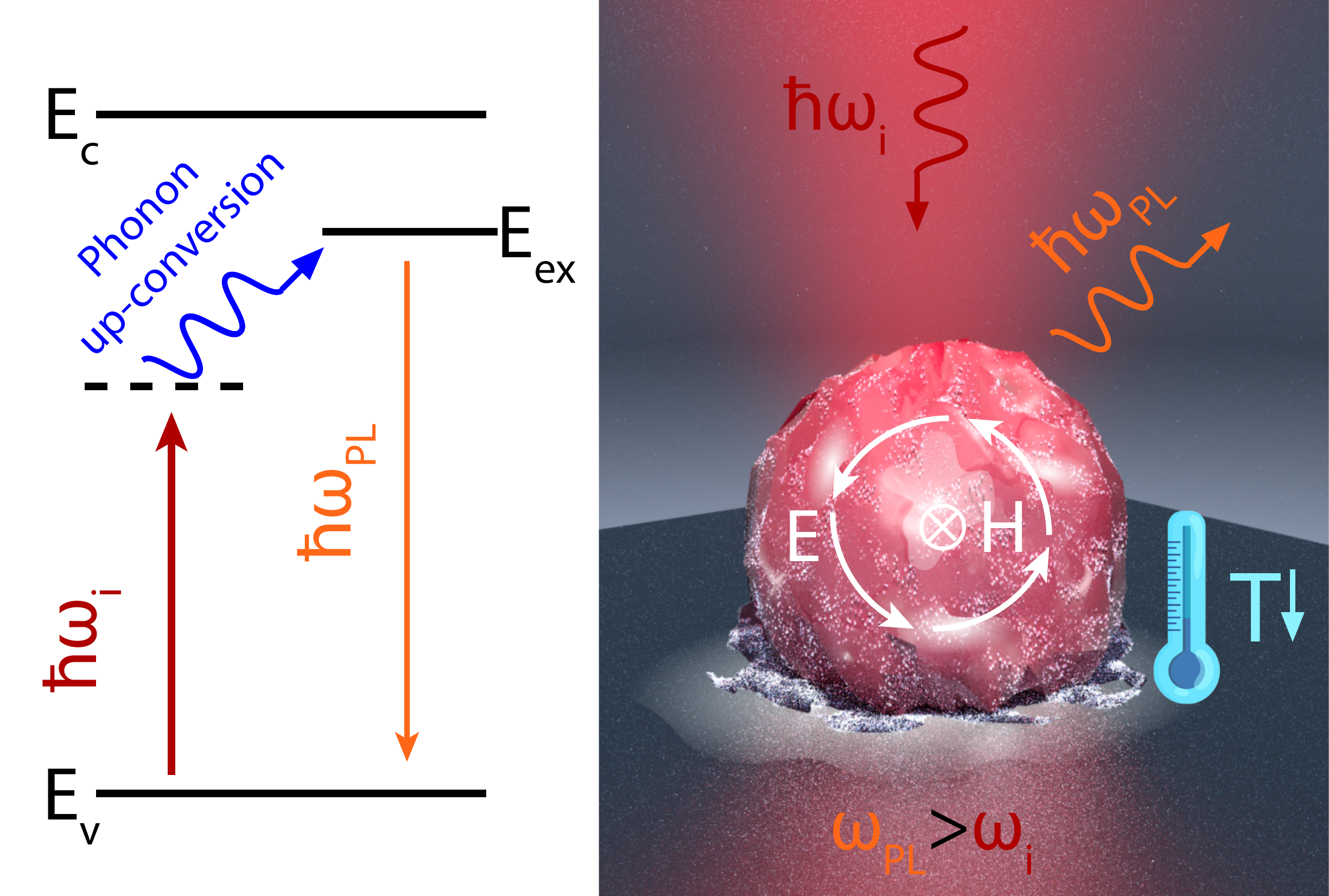}
    \caption{\textbf{Schematic of optical cooling.} The left side depicts a scheme of photoluminescence upconversion mechanism. The right side demonstrates a sketch of optical cooling via enhanced, upon excitation of Mie-modes, upconversion photoluminescence.}
     \label{froz}
\end{figure}

\begin{figure*}[ht!]
    \centering
   
   \includegraphics[width=0.95\linewidth]{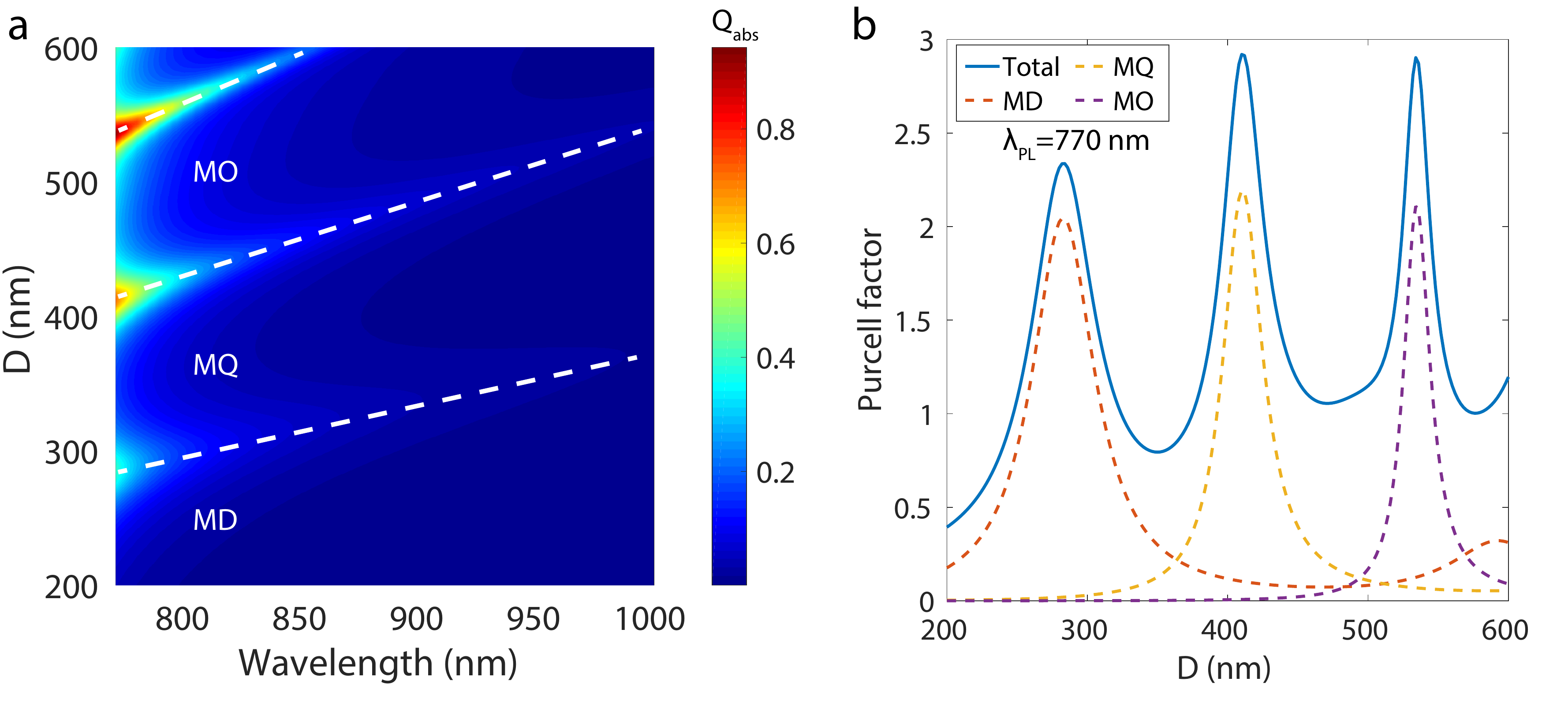}
    \caption{ \textbf{Absorption spectra and Purcell effect in MAPbI$_3$ perovskite NP.} (a) Analytically calculated normalized absorption cross section for spherical MAPbI$_3$ NP with different diameter and excitation wavelength. White dash lines corresponds to Mie-modes: magnetic dipole (MD), magnetic quadrupole (MQ), and magnetic octupole (MO). (b) Analytically calculated Purcell factor for a spherical MAPbI$_3$ NP of different diameters at emission wavelength $\lambda _{PL} $ 770~nm with distinguished contribution of every excited optical mode.}
     \label{ER_Qabs}
\end{figure*}

In this work, we theoretically propose enhanced optical cooling in a resonant perovskite MAPbI$_3$ (MA stands for methylammonium - CH$_3$NH$_3$) NP. By supporting resonant optical modes both at pump wavelength and emission wavelengths, allowing to significantly enhance absorption and radiation efficiency via Purcell effect, the NP demonstrates outstanding performance in laser cooling. As a result, the ability to maintain multiple high-order optical Mie resonances and simple fabrication processes~\cite{makarov2019halide} for the studied material  will pave the way for the nanoscale highly efficient cooling devices.



\section{\label{sec:level1} Results and Discussion}
Fig.~\ref{froz} shows schematially the mechanism of optical cooling of perovskite NP supporting optical resonances both for absorbed and emitted photonos. 



The considered approach of laser cooling via phonon up-conversion photoluminescence, as described by Ha \emph{et al.}~\cite{ha2016laser}, consists of three main steps: i) absorption of the incident pump, ii) generation of the carriers, and iii) emission. Since the ii) is mostly governed by the material and its properties, which was taken the same in this work, i) and iii) can be significantly enhanced by Mie-resonances and Purcell factor~\cite{purcell1946purcell} in the NP, respectively, thus improving the cooling efficiency in total.

\textbf{Photocarrier generation.}
Total carrier concentration  at a laser irradiation is given by sum of equilibrium ($n_0$ for n-type and $p_0$ for p-type) and  photogenerated ($\Delta n$ and $\Delta p$) carrier concentrations:

\begin{align}
    n=n_0+\Delta n \\ 
    p=p_0+\Delta p~.
\end{align}

 The radiation recombination rate ($R_{rad}$) per unit time and per unit volume is proportional to the product of electron and hole concentrations as following~\cite{chuang1995physics}:

\begin{equation}\label{Rrad}
R_{rad}=-Bnp=B(n_0+\Delta n)(p_0+\Delta p)
\end{equation}
where B is  bimolecular recombination
coefficient. 
For lead halide perovskites, the photogenerated carrier concentration is larger than equilibrium carrier concentration~\cite{shi2015low}, i.e. $\Delta n \gg (n_0+p_0)$. In this case, the recombination rate can be evaluated as:
 \begin{equation}\label{RradN}
R_{rad}=BN^2~,
\end{equation}
where $N=\Delta n=\Delta p$ is laser  induced electron-hole carrier density.

For non-radiative recombination, two processes occur: trap-assisted recombination and Auger recombination. The non-radiation recombination rate ($R_{nonrad}$) for this case is given by~\cite{chuang1995physics}:

\begin{equation}\label{RnonradN}
R_{nonrad}=AN+CN^3~,
\end{equation}
where \textit{A} is monomolecular recombination coefficient, \textit{C} is trimolecular recombination coefficient.  

For steady-state conditions with continuum wave (CW)  illumination at a given temperature, the electron-hole carrier density laser induced in bulk material is obtained from the following equation~\cite{sheik2004can}:
\begin{equation}\label{kin1}
\frac{dN}{dt}=\frac{I\alpha}{\hbar\omega_i}-AN-BN^2-CN^3=0~,
\end{equation}
where $I$ is the laser intensity, $\omega_i$ is the frequency of incident light, and $\alpha$ is the absorption coefficient. Therefore, we propose, that the cooling could be significantly increased by enhancing the absorption term $\frac{I\alpha}{\hbar\omega_i}$ via Mie-resonances in a single nanocavity and increasing radiation recombination by boosting spontaneous emission rate ($BN^2$ term). 

\textbf{Absorption.}
Optical response and resonant properties of a nanocavity strongly depend on its permittivity. Since the considered wavelength region for photoluminescence upconversion is below the band gap energy of the MAPbI$_3$ perovskite, the dependence of dielectric permittivity $\varepsilon$ of materials with a strong excitonic contribution can be given by the standard relation for frequency $\omega$: 
\begin{equation}\label{disp}
\varepsilon(\omega)=\varepsilon_0+\frac{f\omega_{exc}^2}{\omega_{exc}^2-\omega^2-i\gamma\omega~,} 
\end{equation}
where $\omega_{exc}$ is the central frequency of excitonic transition, $f$ is the strength of a dipole oscillator, $\gamma$ is the damping factor, $\epsilon_0$ is the background dielectric constant. For MAPbI$_3$ in this work, the parameters values were taken as following~\cite{makarov2019halide}: $\epsilon_0=6.1, \gamma=0.076~eV,  \omega_{exc}=1.68~eV, f=0.02$~(for more details, see Supplementary Information Part I).


Normalized absorption cross-section $C_{abs}$ for spherical NPs is linked with extinction and scattering cross-sections $C_{abs}=C_{ext}-C_{sca}$, where $C_{ext}$ and $C_{sca}$ according to Mie-theory~\cite{bohren2008absorption} with the permittivity taken from Eq.~\ref{disp} were calculated for different diameters of the NP and excitation wavelengths (for details, see Supplementary Information Part II). Results of the calculation are presented in Fig.~\ref{ER_Qabs}a. Three pronounced peaks are observed in the absorption spectra which correspond to magnetic-type modes, namely magnetic dipole (MD) with $\lambda/D=2.67$, magnetic quadrupole (MQ) with $\lambda/D=1.85$ and  magnetic octupole (MO) with $\lambda/D=1.43$, where $\lambda$ is the wavelength in vacuum and $D$ is the NP diameter. Thereby, the enhanced absorption by a single NP is directly related to the number of photogenerated carriers as shown in Eq.~\ref{kin1}.  

\begin{figure}[ht!]
    \centering
   
   \includegraphics[width=1\linewidth]{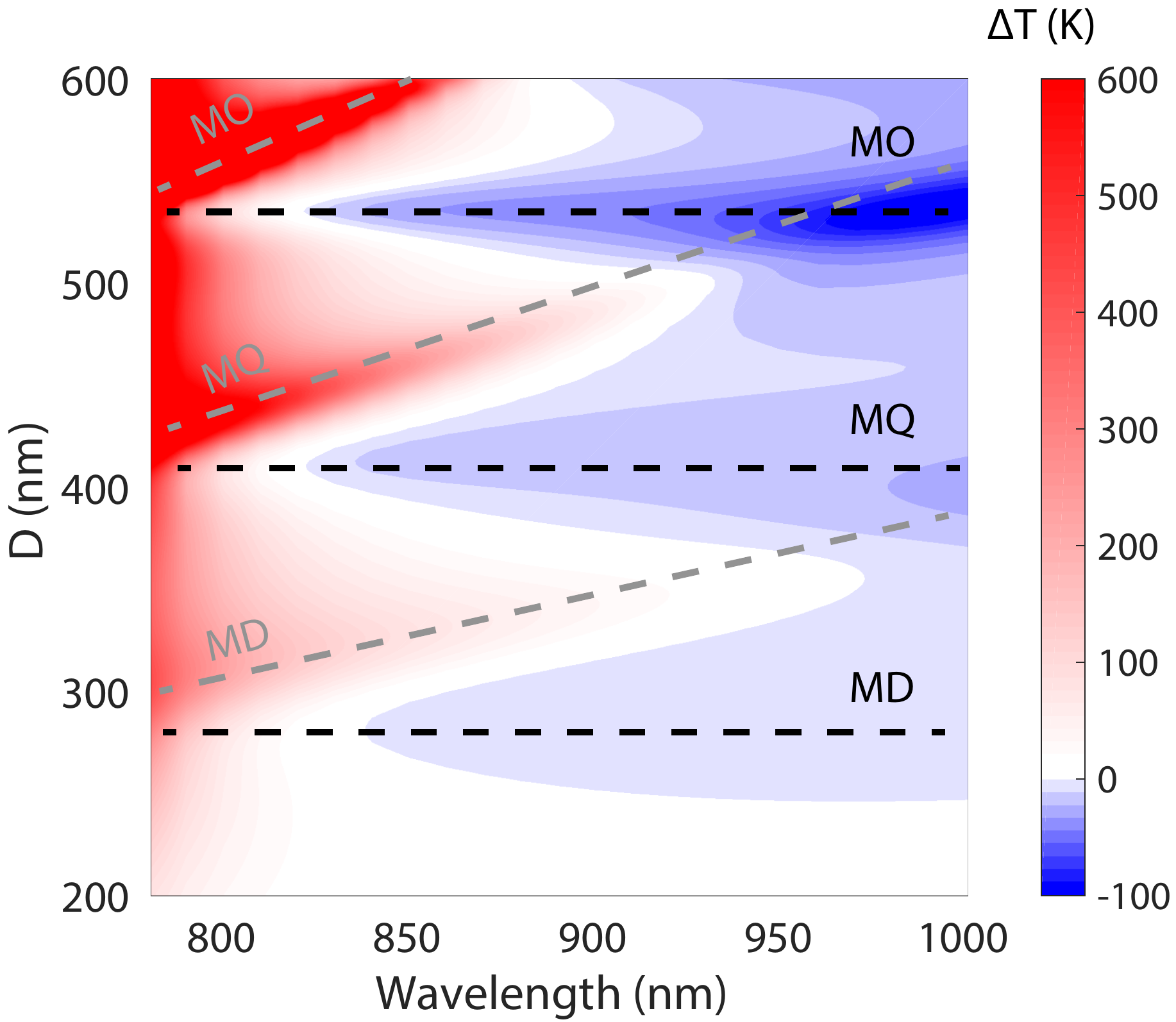}
    \caption{\textbf{Temperature change of a perovskite NP.} Temperature of the spherical MAPbI$_3$ NP with different diameter and pump wavelength $\lambda$ carried out analytically for intensity $I=5~10^5~W/cm^2$ in homogeneous air media using Eq.~\ref{7}. Black dash lines depict modes supported by the NP at emission wavelength $\lambda_{PL} = 770~$nm and gray dash lines show excited modes in absorption.}
    \label{dT} 
\end{figure}

\textbf{Emission rate.}
Another key parameter for the optical cooling is quantum efficiency $\eta$ of emission from the material or nanostructure. Generally, from Eq.~\ref{kin1} one can derive it as following:
\begin{equation}\label{QY}
\eta=\frac{R_{rad}}{R_{rad}+R_{nonrad}},
\end{equation}
Therefore, it is crucial to increase $\eta$ via Purcell effect with a nanocavity. For example, it was shown that for electric dipole transition in a dielectric spherical NP, the average emission rate acceleration can be calculated as described by Chew~\cite{chew1988radiation} (for details, see Supporting Information Part III).









The results for calculation of emission rate enhancement or Purcell factor ($F_p = R_{rad}^{NP}$/$R_{rad}$)
for MAPbI$_3$ NP at wavelength $\lambda$ = 770~nm, which corresponds to the materials fundamental photoluminescence wavelength $\lambda_{PL}$, is presented in Fig.~\ref{ER_Qabs}b (see Supporting Information Part III for details regarding the calculations). Mode decomposition of MAPbI$_3$ NP emission rate demonstrates dominance of the magnetic-type Mie modes similarly to the absorption cross-section. The pronounced peaks correspond to the contributions of 
MD resonance with approximate diameter \textit{D}=283~nm, 
MQ with diameter \textit{D}=410~nm and 
MO with diameter \textit{D}=534~nm at a emission wavelength.

\textbf{Optical cooling in air: an analytical study.}
The ability of nanostructures to support optical cooling is governed by the cooling ratio, which was introduced by~\cite{sheik2004can} as following:  
\begin{equation}\label{Nc}
\eta_c=\eta\frac{\omega_{PL}}{\omega_i}-1~,
\end{equation}
where $\eta$ is luminescence quantum efficiency, $\omega_{PL}$ is PL frequency.
\begin{figure*}[ht!]
    \centering
   
   \includegraphics[width=0.7\linewidth]{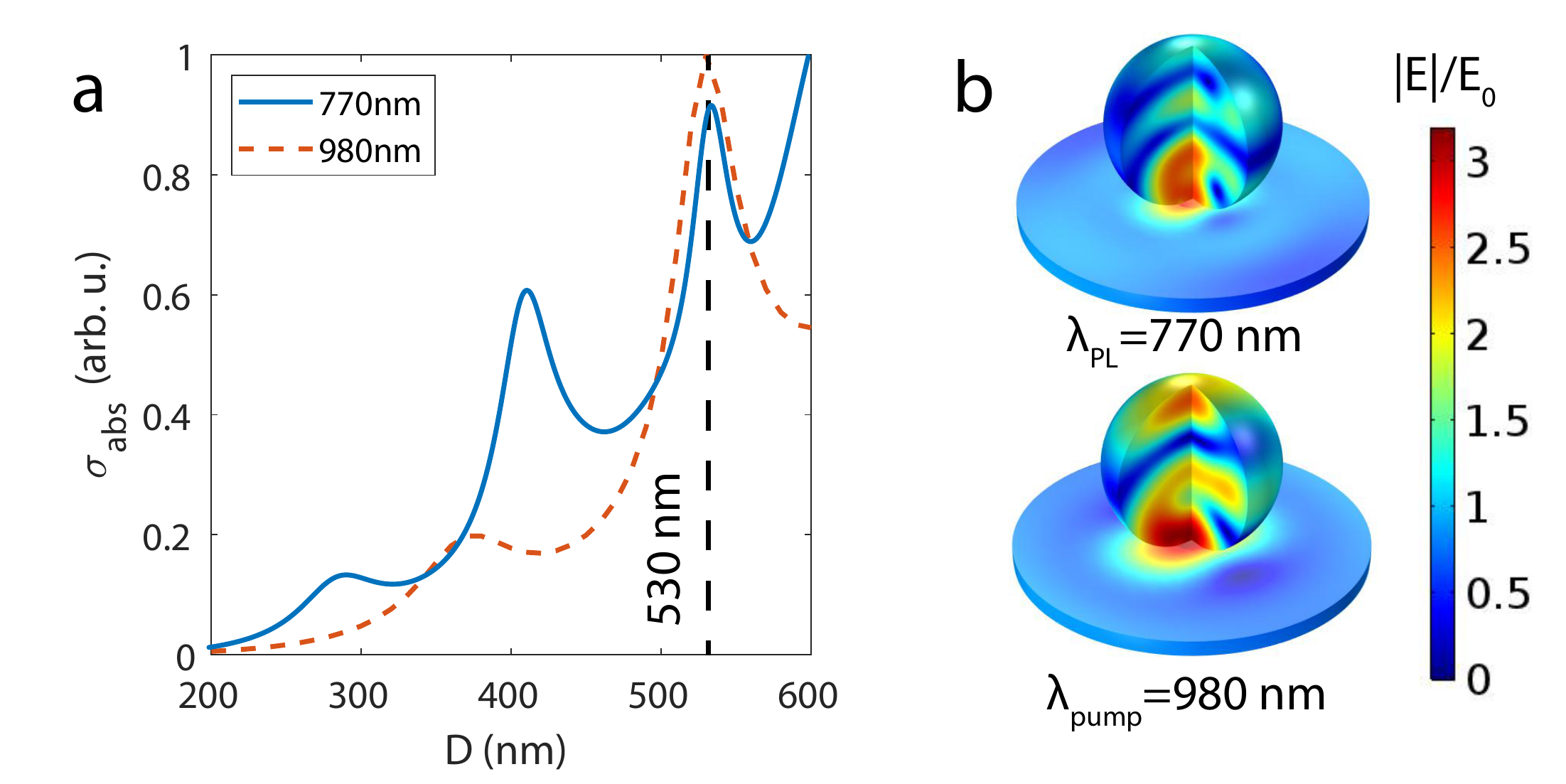}
    \caption{\textbf{Optical properties of a perovskite NP on a glass substrate.} (a) Numerical calculation of the absorption cross section of \(\mathrm{MAPbI_3}\) NP on a glass substrate at pump (red dashed curve) and emission (blue solid curve) wavelengths. (b) Numerical calculation of the electromagnetic field distribution of 530~nm diameter NP at 770~nm and 980~nm wavelengths.}
     \label{Comsol_abs}
\end{figure*}
There are two ways to enhance optical cooling as it follows from Eq.~\ref{Nc}. Firstly, one can increase external radiative quantum efficiency via decreasing the radiation lifetime. As shown in Figure~\ref{ER_Qabs}b, the emission rate undergoes strong resonant enhancement at MD, MQ and MO modes. Secondly, one can decrease frequency of absorbed light by applying the pump sources in the near-IR. As demonstrated in Figure~\ref{ER_Qabs}a, resonant NP can still effectively absorb light in the near-IR at red wing of the excitonic state. Balance of these two effects determines the NP cooling efficiency.

For a single NP, the kinetic equation~\ref{kin1} is modified by taking into account a radiative recombination term and the Mie-enhanced absorption:

\begin{equation}\label{kin2}
\frac{dN}{dt}=\frac{I\sigma_{abs}}{V\hbar\omega}-AN-BF_pN^2-CN^3=0~,
\end{equation}
where $F_p$ is the Purcell factor, $\sigma_{abs}$ is the absorption cross section and $V$ is the volume of the NP.
Solution of Eq.~\ref{kin2} for given nanosphere diameter and wavelength of the excitation gives the value of light induced electron-hole carrier density $N$. 

The cooling efficiency $\eta_c$ is determined by the ratio of the cooling power ($P_{lum}-P_{abs}$) and absorbed power~($P_{abs}$)~\cite{sheik2007optical}:   

\begin{equation}\label{eta_c}
\eta_{c}=\frac{P_{lum}-P_{abs}}{P_{abs}}=\frac{BF_pN^2}{AN+BF_pN^2+CN^3}\frac{\hbar\omega_{PL}}{\hbar\omega_i}-1~, \end{equation}
where the recombination constants ($A=1.4 \; 10^7 s^{-1}$ , $B=9.2 \; 10^{-10} \, cm^{-3}s^{-1}$, $C=1.3 \; 10^{-28} cm^{-6}s^{-1}$) were taken elsewhere~\cite{wehrenfennig2014high}. 

In order to determine the temperature change of the NP under laser illumination one should consider the stationary heat transfer equation and taking into account only radial heat distribution: %
\begin{equation}\label{6}
\nabla(\kappa\nabla T)=\eta_c\frac{I\sigma_{abs}}{V}~,
\end{equation}
where $\kappa$ is thermal conductivity.
Since the thermal conductivity of MAPbI$_3$~\cite{pisoni2014ultra} is much larger than of surrounding air medium~\cite{dean1999lange}, the temperature distribution inside the NP can be considered as homogeneous. The expression for the temperature variance inside the NP can be found by solving the Eq.~\ref{6} and appears to be following (for details, see Supporting Information Part V):
\begin{equation}\label{7}
\Delta T= -\eta_c \frac{\sigma_{abs}I}{2\pi\kappa_{air} D}~,
\end{equation}
where 
$\kappa_{air}$=0.022~$\frac{W}{m\cdot K}$ is the thermal conductivity of the air medium and $D$ is the diameter of a nanosphere.

Fig.\ref{dT} depicts the temperature for different diameters of MAPbI$_3$ NP and excitation wavelengths at moderate laser intensity $I$=5$\cdot$10$^5$ W/cm$^2$ calculated with the Eq.~\ref{7}, where black dashed lines are denoted to the optical modes supported by the sphere at the emission wavelength and gray dashed lines correspond to the optical modes excited in the absorption cross sections. The maximum values of laser cooling are observed at diameter around $D$=530~nm and pump wavelength $\lambda$=980~nm. This matches perfectly with the predictions above. Indeed, NP of diameter $D$=530~nm supports MO resonance at the fundamental MAPbI$_3$ photoluminescence wavelength $\lambda_{PL}$, thus enhancing emission rate, as it is shown in Fig.~\ref{ER_Qabs}. Meanwhile the very same NP possesses MQ resonance nearly the excitation wavelength ($\lambda$=980~nm). Such bi-resonant case of magnetic modes 'interaction' at emission and absorption demonstrates the temperatures decrease down to the lowest values possible. 
Thus, we reveal that the highest temperature decrease is achieved when the nanosphere supports one optical mode at the absorption (incident light) wavelength and the other optical mode at the emission (photoluminescence) wavelength, but both of them are of magnetic nature. According to our simulations, the cooling of larger NPs (D$>$600~nm) is possible, but the value of temperature change is less than that for the most optimal case ($D$=530~nm). 

\begin{figure*}[ht!]\label{dT_980}
    \centering
   
   \includegraphics[width=1\linewidth]{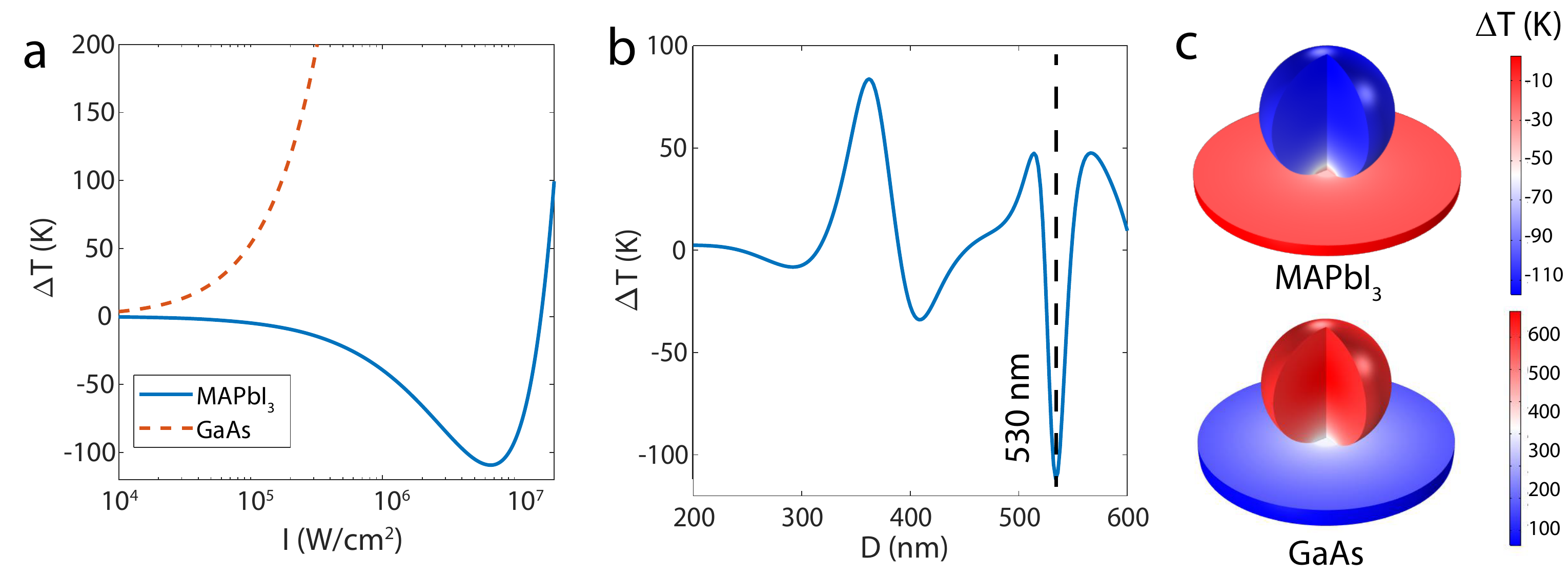}
    \caption{\textbf{Optical cooling of a perovskite NP on a glass substrate.} (a) The dependence of temperature change for 530~nm \(\mathrm{MAPbI_3}\) NP (blue solid line) and 340~nm GaAs NP (red dashed line) on substrate on laser intensity at wavelength $\lambda$=980~nm. (b) The dependence of temperature change for \(\mathrm{MAPbI_3}\) NP on substrate on its diameter upon illumination by laser with intensity \(\mathrm{7\cdot10^{6}W/cm^2}\) and wavelength $\lambda$=980~nm. (c) Temperature distributions for 530~nm \(\mathrm{MAPbI_3}\) NP under 980~nm laser illumination with \(\mathrm{7\cdot10^{6}W/cm^2}\) intensity and 340~nm GaAs NP under laser illumination with the same wavelength and  \(\mathrm{10^{6}W/cm^2}\) intensity.}
    \label{dT_980} 
\end{figure*}

\textbf{Optical cooling on glass: a numerical study.}
In order to prove the possibility of experimental observation of such bi-resonant optical cooling, numerical model was carried out with plane wave irradiation of a \(\mathrm{MAPbI_3}\) NP on a glass substrate~(see Supporting Information Part VII for more details). Fig.~\ref{Comsol_abs}a presents numerical calculation of the normalized absorption cross-section for two wavelengths: 770~nm is the emission wavelength (blue solid line) and 980~nm is the pump wavelength (red dashed line) for \(\mathrm{MAPbI_3}\) nanosphere with different diameters. Elastic optical properties, such as extinction or absorption does not change significantly with the presence of the glass substrate under the NP~\cite{miroshnichenko2015substrate} as well as the spectral position of the resonances (see Supporting Information Part IV for more details). Indeed, as shown in Fig.~\ref{Comsol_abs}, the resonant response remains at very same diameter of $D$=530~nm as it was shown in analytical calculations in Fig.~\ref{dT}. Also, numerically calculated electric field distributions for wavelength of luminescence and pump reveal strong field localization at given wavelengths~(Fig.~\ref{Comsol_abs}b).

The difference between emission rate of the MAPbI$_3$ nanocavity on a glass substrate and in homogeneous air medium is negligible~\cite{zalogina2018purcell,tiguntseva2018light}. For instance, emission rate of 530-nm  MAPbI$_3$ NP at wavelength $\lambda$=770~nm is $F^{air}_p=$2.9, as it follows from the Fig.~\ref{ER_Qabs}, whereas for the same nanosphere in homogeneous glass medium with refractive index 1.45 is $F^{glass}_p$=2.17 (for details, see Supporting Information Part IV), thus we believe, that the presence of a glass substrate from a single bottom side does not affect significantly the resulting emission rate.

According to Eq.~\ref{7}, higher laser intensity at given wavelength and nanocavity parameters leads to enhanced cooling effect. For example, the most optimal design of MAPbI$_3$ NP (\textit{D}=530~nm) yields the lowest temperature ($\Delta T \approx -110K$) at laser intensity $I=7\cdot10^6$~W/cm$^2$ (Fig.~\ref{dT_980}a). As shown in Fig.S3a in Supporting information, this intensity corresponds to photogenerated carriers density $N=3\cdot10^{18}$cm$^{-3}$, where quantum efficiency of luminescence $\eta$ exceeds 86$\%$ level (see Fig.S3b in Supporting information).
At higher intensities, the quantum efficiency decreases owing to increase of Auger recombination rate, resulting in rapid temperature growth. The intensities lower than $I=7\cdot10^6$~W/cm$^2$ can correspond to even slightly better quantum efficiency, but the temperature change directly depends on incident intensity according to Eq.~\ref{7}.




In order to prove that diameter $D$=530~nm of the nanosphere provides the most optimal condition, the dependence of temperature on NP size (Fig~\ref{dT_980}b) was carried out numerically. Indeed, the best cooling from the room temperature of $\Delta T \approx -110$K can be achieved for diameter $D$=530~nm, where both Purcell effect and absorption cross section at magnetic octupole and quadrupole, respectively, are contributed considerably. The steady-state temperature distribution for MAPbI$_3$ nanosphere with 530~nm diameter at 980~nm wavelength excitation with intensity of $I = 7\cdot10^6$W/cm$^2$ and emission at $\lambda$=770~nm is presented in the upper part of Fig.~\ref{dT_980}c. Mostly, the temperature distribution appears to be homogeneous, whereas some temperature gradient occurs in the area of contact with the glass substrate.


In order to compare the obtained results with those for the other materials, let's consider a NP made of another prospective material for optical cooling. Although net cooling has primarily been demonstrated in a Yb$^{3+}$-doped fluoride glass~\cite{epstein1995observation}, the laser cooling of direct-bandgap semiconductors, for example gallium arsenide (GaAs), is more appealing because semiconductors allow for more efficient light absorption, much lower achievable cooling temperatures and direct integrability into electronic and photonic devices~\cite{finkeissen1999cooling, rupper2007optical}.
The lower part of Fig.~\ref{dT_980}c presents the temperature distribution for GaAs nanosphere upon laser irradiation with intensity $I = 10^6$W/cm$^2$. However, the GaAs nanosphere with diameter 340~nm, supporting MQ optical resonant mode at emission wavelength with high Purcell factor, does not undergo any cooling (for more details, see Supplementary Information Part VIII). This happens due to relatively low material inherent quantum yield and non-optimal recombination constants~\cite{t1981radiative, puhlmann1991minority}.

Additionally to the above mentioned advantages, halide perovskites allow for simple fabrication of resonant NPs on arbitrary substrates~\cite{liu2018robust, tiguntseva2018light, gao2018lead}, implying that the parasitic absorption in a substrate material can be eliminated. Moreover, according to the previous works, not only perovskite spheres~\cite{tiguntseva2018light, tang2017single} can be fabricated, but also nanocuboid~\cite{liu2018robust}, cylindrical~\cite{zhizhchenko2019single} and pyramidal~\cite{mi2018fabry} shapes, which have various degrees of freedom for precise manipulation by Mie resonances~\cite{evlyukhin2011multipole, staude2013tailoring}. This opens up unique technological opportunities for creation of highly efficient optical cooling designs.


\section*{Conclusion}
We have demonstrated the resonant phonon-assisted up-conversion photoluminescence optical cooling approach based on enhancement of the emission rate and photoexcitation with Mie resonances at pump and emission wavelengths. The numerical and analytical modeling have revealed that the highest cooling efficiencies for a halide perovskite spherical NP  correspond to the excitation of magnetic-type Mie modes. Namely, magnetic octupole at the emission and magnetic quadrupole at absorption allow for cooling a single nanocavity by $\Delta T$~=~-110~K at realistic conditions. We believe that the proposed mechanism will pave the way for new generation of all-optical nanoscale cooling devices.

\section*{Acknowledgements}
The authors are thankful to Prof. Anvar Zakhidov and Dr. Mihail Petrov for useful discussions. The work was supported by the Ministry of Education and Science of the Russian Federation (Project 14.Y26.31.0010) and Russian Science Foundation (Project 19-73-30023).


\bibliography{ref}

\end{document}